# In silico Identification of tipifarnib-like compounds by structure-based pharmacophore, virtual screening and molecular docking against K-Ras post-translation in colorectal cancer.

Mohammed Mouhcine[1,*], Youness Kadil[1], ImaneRahmoune[1], HoudaFilali[1]

**Corresponding author**

*Correspondence to Mohammed Mouhcine; mouhcine.mohamed@gmail.com.

**Authors, Affiliations**

1. Mohammed Mouhcine;mouhcine.mohamed@gmail.com; [1]Laboratory of Pharmacology-Toxicology, Faculty of Medicine and Pharmacy of Casablanca, Hassan II University of Casablanca, Casablanca, Morocco.
2. Youness Kadil;younesskadil@gmail.com; [1]Laboratory of Pharmacology-Toxicology, Faculty of Medicine and Pharmacy of Casablanca, Hassan II University of Casablanca, Casablanca, Morocco.
3. Imane Rahmoune;rah.imane@gmail.com; [1]Laboratory of Pharmacology-Toxicology, Faculty of Medicine and Pharmacy of Casablanca, Hassan II University of Casablanca, Casablanca, Morocco.
4. Houda Filali;filalihouda@yahoo.fr; [1]Laboratory of Pharmacology-Toxicology, Faculty of Medicine and Pharmacy of Casablanca, Hassan II University of Casablanca, Casablanca, Morocco.

**Abstract**

Colorectal cancer is a public health problem.Approximately 30–50% of colorectal tumors are caused by mutations in the KRAS gene.These mutations induce uncontrolled proliferation.To date,There is no approved effective treatment for the mutated KRAS oncogene.Farnesyltransferase (FTI) inhibitors are considered a therapeutic target against the mutated KRAS oncogene.Tipifarnib is a farnesyltransferase inhibitor that was analyzed in a Phase II trial.In the present study, the three-dimensional structure of farnesyltransferase complexed with tipifarnib [1SA4] was used as a basis to exploit the characteristics of tipifarnib.A pharmacophore model was generated based on the structure using the Asinex (Gold & Platinum Collections) database.A total of 299 molecules were obtained after screening.The 299 molecules were anchored to the tipifarnib binding site in the farnesyltransferase crystal structure for docking analysis.During the molecular docking process, the pharmacophore that was modeled, and was used as a constraint to eliminate the molecules that do not satisfy the pharmacophore.Finally, four Hits identified as farnesyltransferase inhibitors for biological tests.

1. Introduction

Colorectal cancer (CRC) is ranked among the most common cancers in the world and is a major public health problem causing mortality[1][2].About 50% of colorectal adenomas are due to mutated KRAS[3][4][5]. The KRAS gene is a member of the GTPase family[6]. The K-Ras protein oscillates between an active state where it is bound to GTP , and an inactive state where it is bound to GDP [7]. The active state of the K-Ras protein activates the RAS/MAPK signaling pathway and the PI3K/AKT pathway[8][9]. The two RAS/MAPK and PI3K/AKT signaling pathway play a central role in the regulation of cell growth, survival and proliferation[8][10]. Mutations in KRAS gene prevent the interaction of K-Ras with GAPs that hydrolyze the GTP bound to K-Ras. Hence, K-Ras remains in a constitutively active state [11]. This permanent activation of the K-Ras protein will lead to an abnormal permanent activation of mitosis [10][11].

Structurally, the interaction between K-Ras and theireffectorproteinsoccursbetweentwoantiparallel beta strands: β2 of Ras and β2 of RBD(Ras Binding Domain) [12][13].Direct targeting of the Ras-effector interface was difficult due to its relatively flat, shallow and large[14][15]. Due to the difficulty of direct targeting, inhibition of K-Ras protein farnesylation is a therapeutic strategy against colorectal tumors that contain a mutated KRAS gene[16].

The K-Ras protein synthesized in the ribosome immediately undergoes post-translational modifications that facilitate their attachment to the inner surface of the plasma membrane[16][17]. The first and most critical modification is farnesylation by a zinc metalloenzyme protein called farnesyltransferase[16][17]. The process of farnesylation is the transfer of a farnesyl chain from farnesyl pyrophosphate (FPP) into the cysteine residue of certain proteins with a C-terminal CaaX motif (C, cysteine; A, aliphatic amino acid; X, serine or methionine) [18][19][20].Ras isoforms, in general, have CAAX-like C-terminal consensus sequences[20][21] (Figure 1). In such conditions, farnesyltransferase inhibitors (FTI) have been searched to develop specific treatments for tumors with RAS mutations[22].

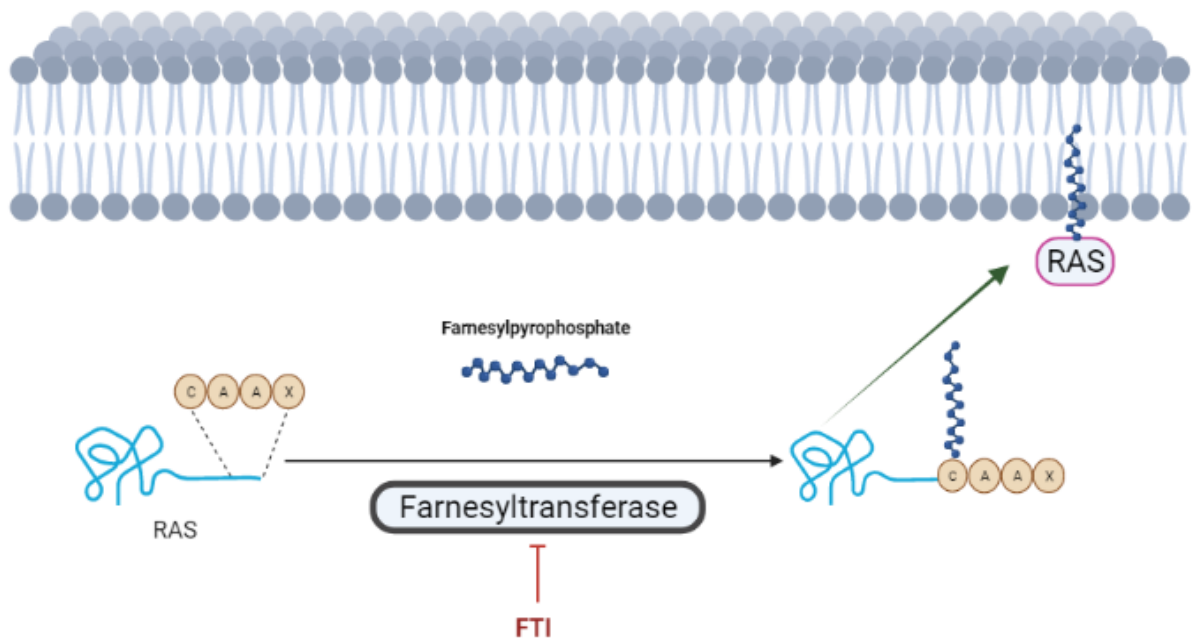

Figure 1: Addition of a FFP on the cysteine of a CAAX motif located in the carboxy terminal of the Ras protein[19].

FTIs are small molecules that can bind at the CAAX binding site, preventing farnesylation and intercalation of the K-Ras protein into the plasma membrane[23][24][25]. Tipifarnib is one of the widely used FTIs as apotentialtreatment of tumors caused by KRAS mutations[26][27].Therefore, there is an urgent need to identify new drugs with similar capabilities to tipifarnib.

The aim of this study is to identify novel FTIs through a computer-aided drug discovery approach. In our study, in silico approaches such as structure-based pharmacophore modeling, structure-based virtual screenings and molecular docking studies were used to identify molecules sharing the same pharmacological activity with tipifarnib.

## 2. Materials and Methods

### 2.1 Structure-Based Pharmacophore Generation

Structure-based pharmacophore modeling was used to identify the critical chemical characteristics of the tipifarnib inhibitor. Therefore, the X-ray crystallographic structure of farnesyl transferase in complex with Tipifarnib at high resolution was obtained from the Protein Data Bank (PDB ID :1SA4) To construct a reliable pharmacophore model based on the structure[28]. In this structure tipifinar is competitive with farnesyl diphosphate. A molecular interaction diagram of tipifarnib between farnesyl transferase and FFP was generated using Ligand Interactions application of MOE(Molecular Operating Environment) to determine the key interaction points of the inhibitor Tipifarnib. The chemical characteristics of Tipifarnib were therefore created using the Pharmacophore Query Editor module of MOE . On the basis of interaction diagram a pharmacophore with aromatic ring ,hydrogen bond acceptor and hydrophobic features.

## 2.2 Virtual Screening of the Asinex (Gold & Platinum) Collections Database To Retrieve the Potential Compounds

the obtained pharmacophore model was used as a 3D query to screen the Asinex (Gold & Platinum Collections) database, which contains 261120 drug-like molecules[29]**.**Virtual screening was used by the Pharmacophore Search module in the MOE software. The purpose of the Pharmacophore Search is to use a pharmacophore query to filter databases of 3D conformations based on the positions of annotation points derived from each of the ligand conformations.During the screening process, a conformational search was performed on the candidate molecules before launching the search. Then, the research is applied to these conformations. Candidate ligands that match the pharmacophore model were selected for further molecular docking analysis.

## 2.3 Structure-Based Molecular Docking

In-silico molecular docking is a method that predicts the interactions and molecular orientationsbetween ligands and receptors. The MOE Dock module was used to perform the molecular dockingstudies. In the present work, the results of the virtual screening were docked to the active site offarnesyl transferase using the default triangular matching algorithm. The free energy of bindingbetween farnesyltransferase and a ligand was estimated by The London dG docking evaluation function of MOE (lower values indicate better binding affinity). The obtained pharmacophore is usedas a constraint to guide the docking process. All final poses that do not satisfy the pharmacophore (inabsolute mode) will be eliminated. The final results were selected as farnesyltransferase inhibitorsand similar with tipifarnib.

## Results and Discussion

## Examining the Binding Mode of tipifarnib

Tipifarnib atoms can form hydrogen bonds with solvent molecules, usually water (Figure 2). Two possible hydrogen (H-donor) interactions between tipifarnib and farnesyl diphosphat, results showed the initial interaction with a distance of 3.70 Å and an energy of -1.1 Kcal/mol between the tipifarnib c1 atom and the o2A atom of farnesyl diphosphat, a second interaction was given with a distance of 3.74 Å and an energy of -0.7 Kcal/mol between the tipifarnib c1 atom and the o3A atom of farnesyl diphosphate (figure 2 A,B).

Two possible pi-H interactions between tipifarnib and farnesyltransferase through water molecules are: An interaction with a distance of 3.92 Å and an energy of -2.4 Kcal/mol between 5-ring of tipifarnib and the o-atom of H2O; Other interaction with a distance of 3.95 Å and an energy of -3.2 Kcal/mol between 6-ring of tipifarnib and the o-atom of H2O(figure c).

A metal interaction with a distance of 2.10 Å and an energy of -4.6 Kcal/mol between the zinc ion and the N2 atom of tipifarnib (figure 2). Based on these possible interactions a pharmacophore was generated for virtual screening.

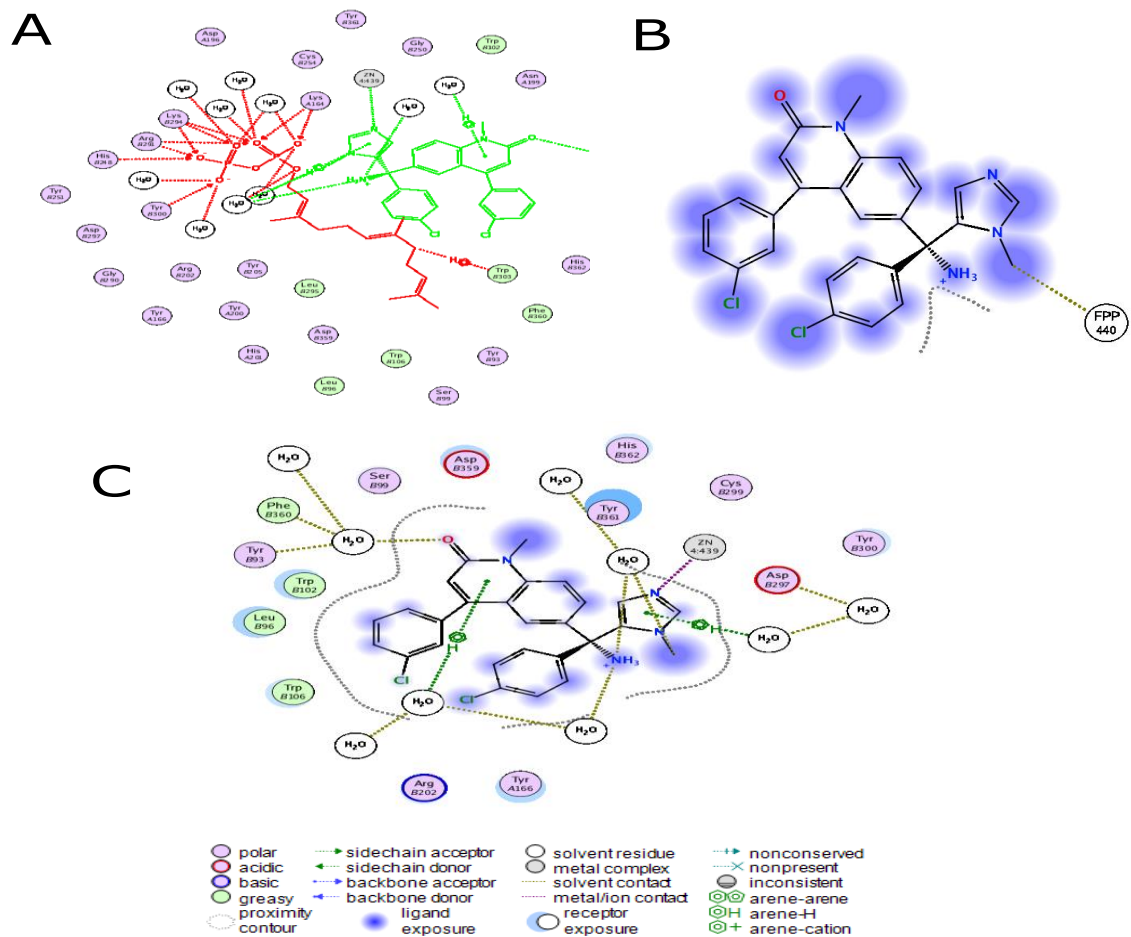

Figure 2: Interaction diagram between tipifarnib and farnesyl diphosphate (A,B). Interaction diagram between tipifarnib and farnesyltransferase (C ).

**Pharmacophore Modeling**

After obtaining all available chemical information on the binding of the inhibitor tipifarnib with farnesyltransferase a structure-based pharmacophore model was constructed (Figure 3). The model was composed of six features (Figure 3A): two hydrogen bond acceptor elements (F3 and F5:Acc), three aromatic elements (F1,F2,F4:Aro) and two hydrophobic elements (F4 and F6:Hyd). These features represent the key interaction points for binding of the inhibitor tipifarnib to farnesyltransferase : (i) two features Acc and Aro (F3 and F1) corresponding to

Tyr300, Asp297, zinc ion and water molecules; (ii) two characteristics Hyd|Aro and Hyd (F6 ef F4) corresponding to farnesyl diphospha (FPP440); (iii) two features Aro and Acc(F2 and F5) corresponding to Asp359,His362,Tyr361 and water molecules (Figure 3).

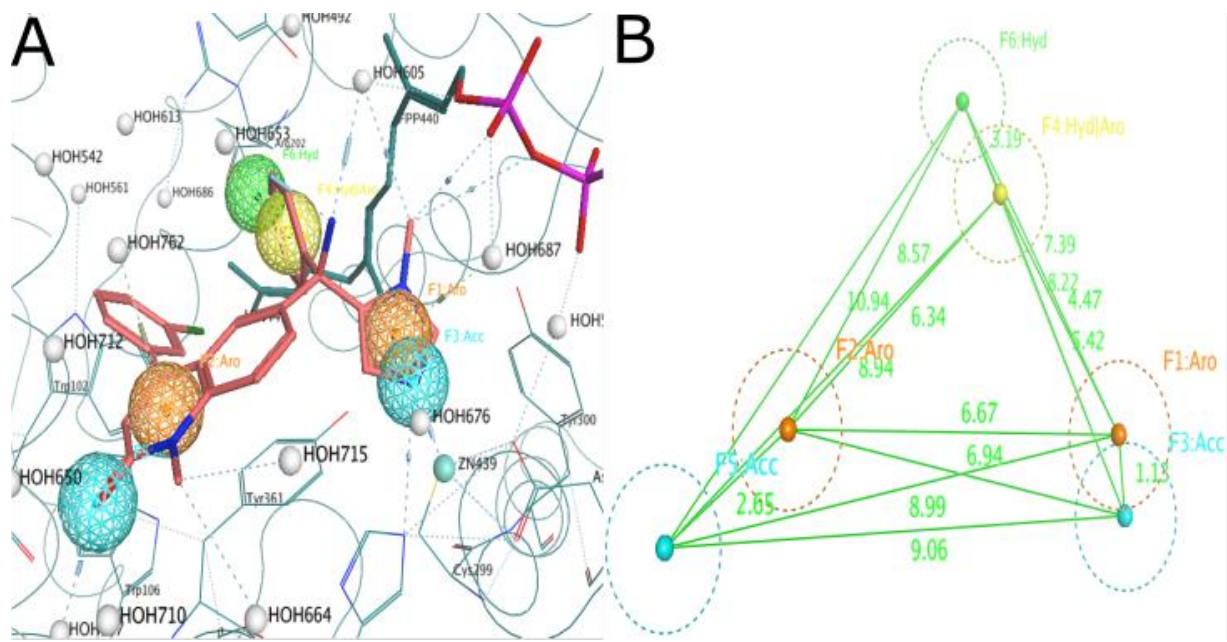

Figure 3. Generation of a pharmacophore model exploiting Tipifarnib. (A) Key features demonstrated by Tipifarnib. (B) The distances between features.

**Pharmacophore-based virtual screening & structure-Based Molecular Docking**

Pharmacophore-based screening of a compound database was used to identify new inhibitors of the protein farnesyltransferase. The different phases of this strategy are presented in Figure 4.

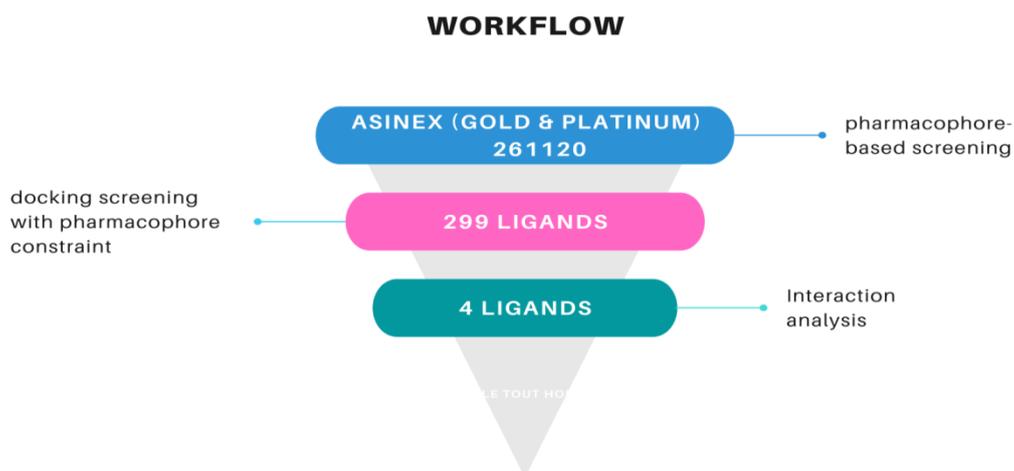

Figure 4. Different stages of database screening.

First, the 261120 molecules from the Asinex database (Gold & Platinum) were screened against the pharmacophore model. A total of 299 molecules were selected, The 299 molecules were anchored to the tipifarnib binding site in the farnesyltransferase crystal structure for docking analysis. the pharmacophore that was modeled in our study is used as a constraint to eliminate molecules that do not satisfy the pharmacophore. Consequently, four compounds, called hits 1-4, were finally selected for further biological evaluation later (Table 1).RMSDwas used as a distance measure between the query features and the corresponding ligand annotation points. The RMSD of the four hits had a value less than or equal to 0.757Å, which indicates a good pharmacophore mapping of 4 hits on the model. The purpose of docking is to anticipate the possibility that any ligand may or may not lodge in a protein pocket with good affinity[30]. From Table 1,The free energy of binding of the four hits with the active site of the farnesyltransferase protein ranged from -6.68 kcal/mol to -7.93 kcal/mol indicate a better binding affinity of the four hits. Figure5 shows that the 4 hits share the same chemical characteristics of the pharmacophore model.

The table represent the chemical structures of different compounds selected from asinex database, wich can have a like effect of Tipifarnib (Table 1).

Table 1: Compounds selected from the Asinex database (Gold & Platinum). RMSD: The root of the mean square distance between the pharmacophore points of the molecule and the pharmacophore points of the model. Docking Score: free energy of binding between the protein farnesyltransferase and a PubChem CID: molecule identifier in the PubChem chemical molecules database.

| Hits | Chemical structure | RMSD(Å) | Docking Score (kcal·mol−1) | ID PubChem |
|---|---|---|---|---|
| 1 | 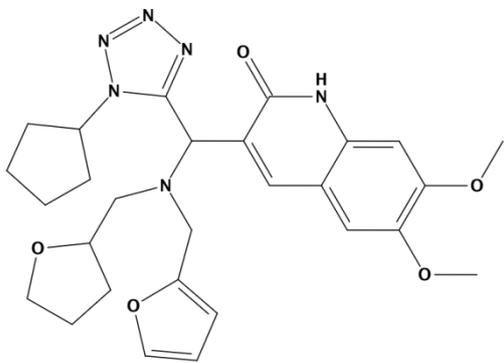 | 0.757 | -7.93 | 3187875 |
| 2 | 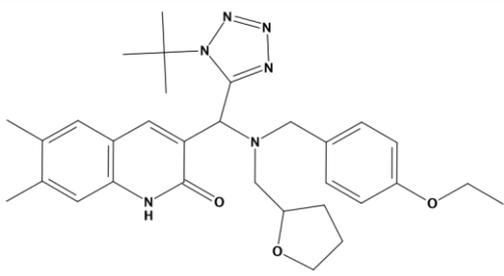 | 0.523 | -7.40 | 3187691 |
| 3 | 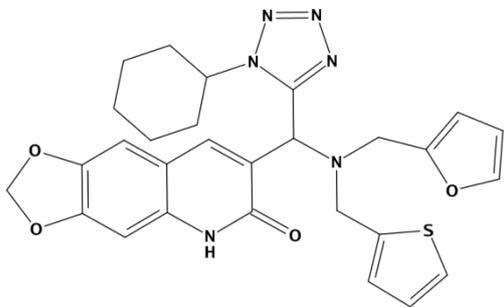 | 0.626 | -7.13 | 3150518 |

| 4 | 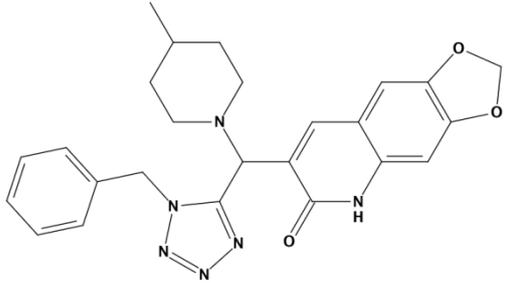 | 0.704 | -6.68 | 3183757 |

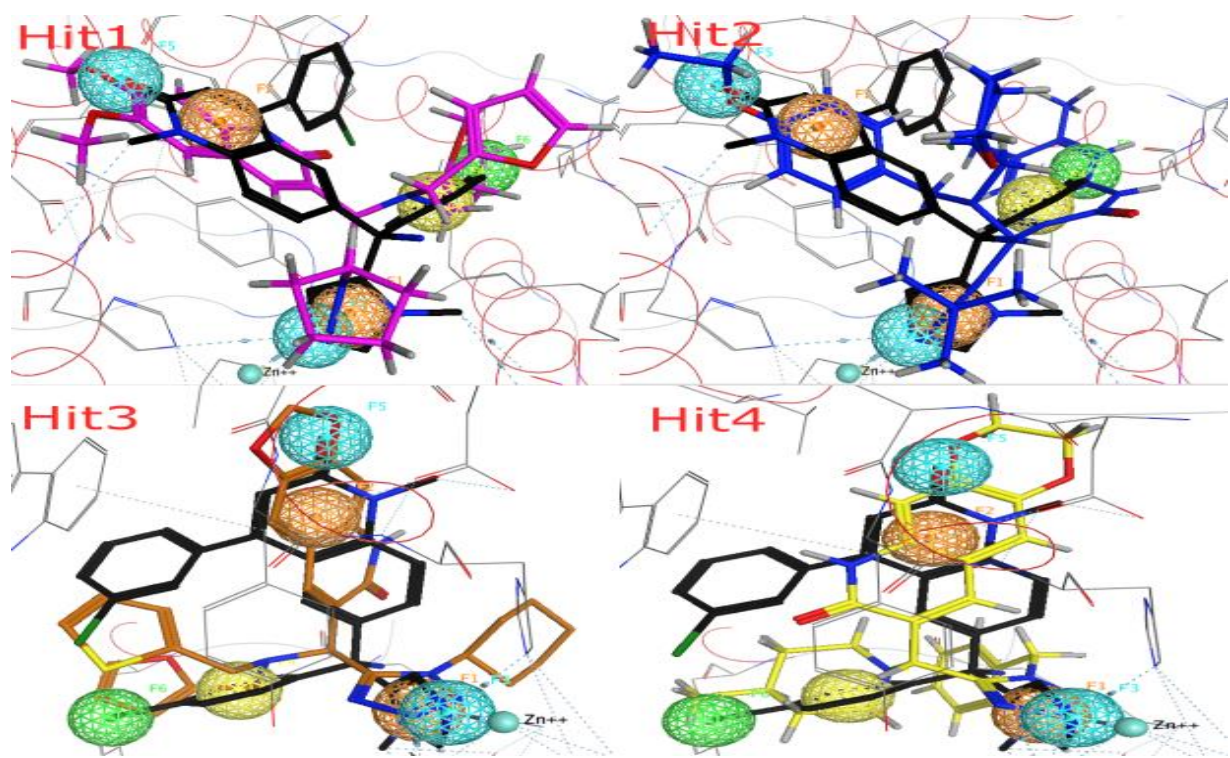

Figure 5 : Pharmacophore mapping of four hits and tipifarnib on model. F3 and F5 are H-bond acceptor features colored with cyan color,F4 is aromatic or hydrophobic feature colored a yellow color,F1 and F2 are aromatic features colored colored with a brown color,F6 is ahydrophobic feature feature colored with a green color.The hits are overlapping with tipifarnib which colored in black.protein backbone shown in line form.

## Molecular Docking Study

The different interaction of the four selected compounds with farnesyltransferase are provided in Table 2.

Table 2: key interactions between the hits and farnesyltransferase.

| Hits | Hydrogen Bond type | van der Waals | ion interaction |
|------|--------------------|----------------|------------------|
| Hit1 | Cys299(H-acceptor) Tyr361(π-π) | Asp352, Asp297, Tyr300, His201, Asn165, Arg202, phe360, Lys164, Tyr166, His362, Asp359, Trp106, Tyr93, Leu96, Leu103 | _ |
| Hit2 | Cys299(H-donor) Cys299(H-acceptor) | Tyr300, Asp297, Asp352, Tyr166, Trp106, Leu96, Tyr93, Tyr359, Asp359, His362 | zinc |
| Hit3 | Cys299(H-acceptor) Tyr166(H-pi) | Asp297, Phe360, His362, Tyr361, Lys164, Asp359, Leu96, Tyr93, Trp102, Trp106, Arg202 | zinc |
| Hit4 | Tyr361 (π-π) Cys299(H-acceptor) | Lys164, Tyr166, Arg202, Trp106, His362, Asp359, Tyr93, Leu96, Tyr300, Asp297, His201 | - |

All the 2D interaction diagram presentation of the four hits are displayed in figure 5.

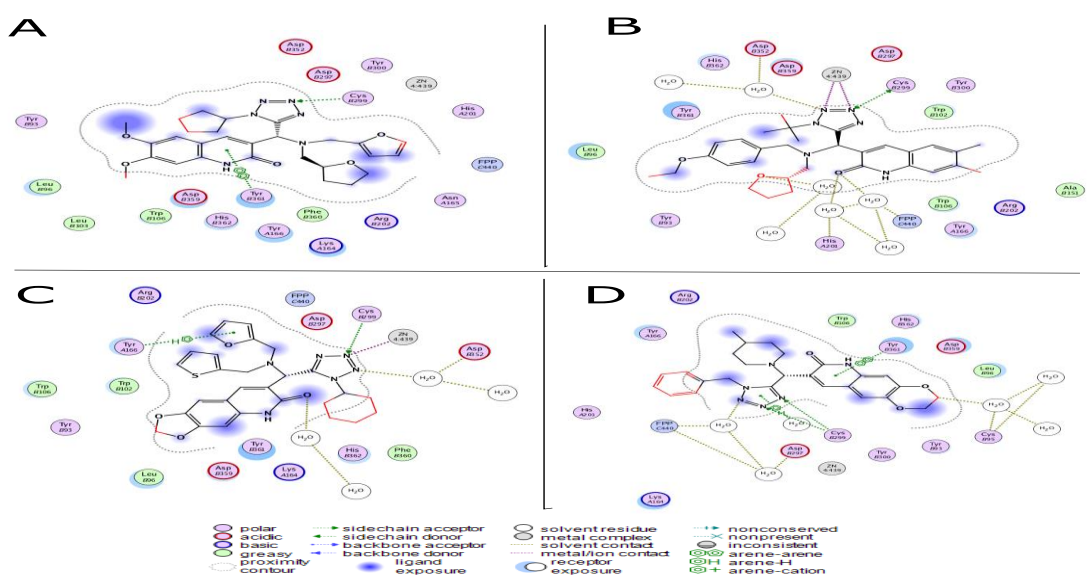

Figure 5: Binding interaction of compounds (A) Hit1, (B) Hit2, (C) Hit3, (D) Hit4 in the active site of the protein human farnesyltransferase.

Figure 5 shows different interactions were found between farnesyltransferase and the four hits, such as hydrogen interactions and van der wals interactions. According to the result of molecular docking, The four hits interact with the residues of the active site of fransyltransferase. Furthermore, hit1 and hit2 were interacted with zinc ion like tipifarnib. Furthermore, the four hits identified have different scaffolds and represent the main features of the pharmacophore model. These four hits may outperform tipifarnib in pharmacokinetic and pharmacodynamic tests. According to the results obtained, these four compounds may be new drugs against colorectal cancer.

## Conclusions

On the basis of our results of the virtual pharmacophore-based screening, we can conclude that the four obtained and identified compounds may act as a potential agentsagainst colorectaltumurs induced by KRAS mutation, involving farnesyltransferase.

## Acknowledgements
None.

## Funding
None.

## Authors' contributions

All authors listed have made a substantial, direct and intellectual contribution to the work, and approved it for publication.

## Declarations

### Ethics approval and consent to participate

None.

### Consent for publication

None.

**Competing interests**

The authors declares that there is no conflict of interest.

**References**

[1]   I. Mármol, C. Sánchez-de-Diego, A. P. Dieste, E. Cerrada, and M. J. R. Yoldi, "Colorectal carcinoma: A general overview and future perspectives in colorectal cancer," *Int. J. Mol. Sci.*, vol. 18, no. 1, 2017, doi: 10.3390/ijms18010197.

[2]   A. Nasrallah and M. El-Sibai, "Colorectal cancer causes and treatments: A minireview," *Open Color. Cancer J.*, vol. 7, no. 1, pp. 1–4, 2014, doi: 10.2174/1876820201407010001.

[3]   D. Uprety and A. A. Adjei, "KRAS: From undruggable to a druggable Cancer Target," *Cancer Treat. Rev.*, vol. 89, no. July, p. 102070, 2020, doi: 10.1016/j.ctrv.2020.102070.

[4]   P. Liu, Y. Wang, and X. Li, "Targeting the untargetable KRAS in cancer therapy," *Acta Pharm. Sin. B*, vol. 9, no. 5, pp. 871–879, 2019, doi: 10.1016/j.apsb.2019.03.002.

[5]   D. S. Bhullar, J. Barriuso, S. Mullamitha, M. P. Saunders, S. T. O'Dwyer, and O. Aziz, "Biomarker concordance between primary colorectal cancer and its metastases," *EBioMedicine*, vol. 40, pp. 363–374, 2019, doi: 10.1016/j.ebiom.2019.01.050.

[6]   C. H. Gray, J. Konczal, M. Mezna, S. Ismail, J. Bower, and M. Drysdale, "A fully automated procedure for the parallel, multidimensional purification and nucleotide loading of the human GTPases KRas, Rac1 and RalB," *Protein Expr. Purif.*, vol. 132, pp. 75–84, 2017, doi: 10.1016/j.pep.2017.01.010.

[7]   S. I. ABRAMS, "Development of Epitope-Specific Immunotherapies for Human Malignancies and Premalignant Lesions Expressing Mutated ras Genes," *Gene Ther. Cancer*, pp. 145–163, 2002, doi: 10.1016/b978-012437551-2/50009-4.

[8]   R. Nussinov, C. J. Tsai, and H. Jang, "Does Ras Activate Raf and PI3K Allosterically?," *Front. Oncol.*, vol. 9, no. November, pp. 1–10, 2019, doi:

10.3389/fonc.2019.01231.

[9] X. Z. S. Xiang, W. Bai, G. Bepler, *Conquering RAS*, Academic P. 2017.

[10] Y. Aoki, T. Niihori, Y. Narumi, S. Kure, and Y. Matsubara, "The RAS/MAPK syndromes: Novel roles of the RAS pathway in human genetic disorders," *Hum. Mutat.*, vol. 29, no. 8, pp. 992–1006, 2008, doi: 10.1002/humu.20748.

[11] R. Dolatkhah, M. H. Somi, R. Shabanloei, F. Farassati, A. Fakhari, and S. Dastgiri, "Main risk factors association with proto-oncogene mutations in colorectal cancer," *Asian Pacific J. Cancer Prev.*, vol. 19, no. 8, pp. 2183–2190, 2018, doi: 10.22034/APJCP.2018.19.8.2183.

[12] F. M. & A. W. Nicolas Nassar, Gudrun Horn, Christian A. Herrmann, Anna Scherer, *The 2.2 Å crystal structure of the Ras-binding domain of the serine/threonine kinase c-Raf1 in complex with RaplA and a GTP analogue.* 1995.

[13] S. Lu *et al.*, "Drugging Ras GTPase: A comprehensive mechanistic and signaling structural view," vol. 45, no. 18, pp. 4929–4952, 2017, doi: 10.1039/c5cs00911a.Drugging.

[14] A. Ernst, "Competitive Inhibitors of Ras Effector Binding," pp. 1–47, 2019.

[15] M. Nagasaka, B. Potugari, A. Nguyen, A. Sukari, A. S. Azmi, and S. H. I. Ou, "KRAS Inhibitors– yes but what next? Direct targeting of KRAS– vaccines, adoptive T cell therapy and beyond," *Cancer Treat. Rev.*, vol. 101, no. August, p. 102309, 2021, doi: 10.1016/j.ctrv.2021.102309.

[16] S. Gysin, M. Salt, A. Young, and F. McCormick, "Therapeutic strategies for targeting Ras proteins," *Genes and Cancer*, vol. 2, no. 3, pp. 359–372, 2011, doi: 10.1177/1947601911412376.

[17] C. J. M. John F. Hancock, Hugh Paterson, *A polybasic domain or palmitoylation is required in addition to the CAAX motif to localize p21ras to the plasma membrane*. Cell, 1990.

[18] K. M. M. M.A. Avery, D.J. Weldon, "Drugs for Parasitic Infections: Advances in the Discovery of New Antimalarials," 2016, doi: https://doi.org/10.1016/B978-0-12-409547-2.11056-X.

[19]  C. B. Costa *et al.*, "Farnesyltransferase inhibitors: Molecular evidence of therapeutic efficacy in acute lymphoblastic leukemia through cyclin D1 inhibition," *Anticancer Res.*, vol. 32, no. 3, pp. 831–838, 2012.

[20]  S. F. Sousa, P. A. Fernandes, and M. J. Ramos, "Farnesyltransferase - New insights into the zinc-coordination sphere paradigm: Evidence for a carboxylate-shift mechanism," *Biophys. J.*, vol. 88, no. 1, pp. 483–494, 2005, doi: 10.1529/biophysj.104.048207.

[21]  N. S. H. N. Moorthy, S. F. Sousa, M. J. Ramos, and P. A. Fernandes, "Molecular dynamic simulations and structure-based pharmacophore development for farnesyltransferase inhibitors discovery," *J. Enzyme Inhib. Med. Chem.*, vol. 31, no. 6, pp. 1428–1442, 2016, doi: 10.3109/14756366.2016.1144593.

[22]  S. Rao *et al.*, "Phase III double-blind placebo-controlled study of farnesyl transferase inhibitor R115777 in patients with refractory advanced colorectal cancer," *J. Clin. Oncol.*, vol. 22, no. 19, pp. 3950–3957, 2004, doi: 10.1200/JCO.2004.10.037.

[23]  A. Kazi *et al.*, "Dual farnesyl and geranylgeranyl transferase inhibitor thwarts mutant KRAS-driven patient-derived pancreatic tumors," *Clin. Cancer Res.*, vol. 25, no. 19, pp. 5984–5996, 2019, doi: 10.1158/1078-0432.CCR-18-3399.

[24]  C. J. Novotny *et al.*, "Farnesyltransferase-mediated Delivery of a Covalent Inhibitor Overcomes Alternative Prenylation to Mislocalize K-Ras," vol. 12, no. 7, pp. 1956–1962, 2018, doi: 10.1021/acschembio.7b00374.Farnesyltransferase-mediated.

[25]  M. Baranyi, L. Buday, and B. Hegedűs, "K-Ras prenylation as a potential anticancer target," *Cancer Metastasis Rev.*, vol. 39, no. 4, pp. 1127–1141, 2020, doi: 10.1007/s10555-020-09902-w.

[26]  P. Obasaju, K. Pollard, A. Allen, J. Wang, L. Kessler, and C. A. Pratilas, "Inhibition of farnesyl transferase by tipifarnib leads to cell growth inhibition in HRAS-mutated human rhabdomyosarcoma," *Eur. J. Cancer*, vol. 138, p. S46, 2020, doi: 10.1016/S0959-8049(20)31199-0.

[27]  J. V. Heymach *et al.*, "Phase II study of the farnesyl transferase inhibitor R115777 in patients with sensitive relapse small-cell lung cancer," *Ann. Oncol.*, vol. 15, no. 8, pp. 1187–1193, 2004, doi: 10.1093/annonc/mdh315.

[28] T. S. R. and L. S. Beese, "Crystal Structures of the Anticancer Clinical Candidates R115777 (Tipifarnib) and BMS-214662 Complexed with Protein Farnesyltransferase Suggest a Mechanism of FTI Selectivity†," *Biochemistry*, 2004, doi: https://doi.org/10.1021/bi049723b.

[29] "Asinex. [(accessed on 29 August 2016)]." [Online]. Available: http://www.asinex.com.

[30] R. M. Lengauer T, "Computational methods for biomolecular docking," *Curr. Opin. Struct. Biol.*, vol. 6, no. 3, 1996, doi: 10.1016/s0959-440x(96)80061-3.